\documentclass[preprintnumbers,superscriptaddress]{revtex4}

\usepackage{amssymb}
\usepackage{epsfig,rotating}

\begin{document}

\title{Central Diffractive Processes at the Tevatron, RHIC and LHC\footnote{Talk given by V.A. Khoze at International Workshop on
Diffraction in High-Energy Physics, Otranto (Lecce), Italy,
September 10 - 15, 2010.}
}




\author{L.A. Harland-Lang}
\affiliation{Cavendish Laboratory, University of Cambridge, J.J. Thomson Avenue, Cambridge, CB3 0HE, UK}
\author{V.A. Khoze}
\affiliation{Department of Physics and Institute for Particle Physics Phenomenology, University of Durham, DH1 3LE, UK}
\affiliation{School of Physics \& Astronomy, University of Manchester, Manchester M13 9PL, UK}
\author{M.G. Ryskin}
\affiliation{Department of Physics and Institute for Particle Physics Phenomenology, University of Durham, DH1 3LE, UK}
\affiliation{Petersburg Nuclear Physics Institute, Gatchina, St. Petersburg, 188300, Russia}
\author{W.J. Stirling}
\affiliation{Cavendish Laboratory, University of Cambridge, J.J. Thomson Avenue, Cambridge, CB3 0HE, UK}
\affiliation{Department of Physics and Institute for Particle Physics Phenomenology, University of Durham, DH1 3LE, UK}

\preprint{IPPP/10/93}
\preprint{DCPT/10/186}
\preprint{Cavendish-HEP-10/20}

\begin{abstract}
Central exclusive production (CEP) processes in high-energy hadron collisions offer a very promising framework for studying both novel aspects of QCD and new physics signals. We report on the results of a theoretical study  of the CEP of heavy quarkonia ($\chi$ and $\eta$) at the Tevatron, RHIC and LHC (see \cite{HarlandLang09}~--~\cite{khrysthal} for details). These processes  provide important information on the physics of bound states and can probe the current ideas and methods of QCD, such as effective field theories and lattice QCD.
\end{abstract}

\maketitle

Recently there has been a renewal of interest in studies of CEP processes in 
high-energy proton -- (anti)proton collisions,
\cite{KMRprosp}~--~\cite{Royon}. In particular, such measurements represent
 a promising  way to study the properties of new particles, from exotic hadrons to the Higgs boson
(e.g. \cite{HarlandLang09}~--~\cite{khrysthal}, \cite{Khoze04}~--~\cite{teryaev}).
 The CEP of an object $X$ may be written in the form 
\begin{equation}\nonumber
pp({\bar p}) \to p+X+p({\bar p})\;,
\end{equation}
where $+$ signs are used to denote the presence of large rapidity gaps. An attractive advantage of these reactions is that they provide an especially clean environment in which to measure the nature and quantum numbers (in particular, the spin and parity) of new states, see for instance~\cite{Khoze00a,kkmr}.
An important example is the CEP of the Higgs boson~\cite{acf,HKRSTW,ways}, which provides a novel route to study in detail the Higgs sector at the LHC and is complementary to the conventional production mechanisms \cite{ways,fp420}.

CEP processes have been successfully observed  at the Tevatron
\cite{MA,Royon} by selecting events 
with large rapidity gaps separating the centrally produced state from the dissociation products of incoming protons. 
The CDF measurement \cite{cdfchic}
of $\chi_c$ CEP, for which the experimental signature is especially well-defined, is of particular interest: $p + \bar{p} \rightarrow p + \chi_c + \bar{p}$ with \emph{no other
particles} in the final state.
The CDF result, 
$d\sigma/dy|_{y=0} = 76\pm10(\mathrm{stat})\pm 10(\mathrm{syst})$ nb in the $\chi_c \rightarrow J/\psi+\gamma$ channel with
$J/\psi \rightarrow \mu^+\mu^-$, was in reasonable agreement with the 
earlier {\it pre}diction by Durham group  \cite {Khoze04} (see also \cite{HarlandLang09}). 
The broad agreement of the Tevatron results on all CEP processes with
the theory lends credence to the Durham theoretical framework and
 motivates further investigation of new and SM CEP physics at the Tevatron, LHC and RHIC.

The central diffractive production programme at the LHC looks very promising, and  the
first ALICE results were reported in \cite{schciker}.
However, currently all LHC experiments have insufficient
forward coverage, which does not allow a full reconstruction of CEP processes.
 As emphasized at this meeting by Mike Albrow \cite{MA} and Risto Orava \cite{RO}, the
{\it uninstrumented} rapidity
 gaps can be covered 
with sets of simple scintillation counters (FSC = Forward Shower Counters)~\cite{fsc} along the beam pipes. 
This will allow the selection of $p+X+p$ events,
without actually detecting the protons (which  could  be done with the TOTEM and ALFA detectors).
ALICE is installing such counters and they are proposed for CMS \cite{MA}.

A new area of experimental studies of CEP with {\it tagged} forward protons
is now being explored by the STAR Collaboration at RHIC \cite{wlodek}, which
has the capability to trigger on and to measure the outgoing forward protons,
providing at the same time measurement of the central system with excellent mass resolution. 
In \cite{khrysthal} we pay special attention to exclusive charmonium
($\chi_{cJ}$ and $\eta_c$) production at RHIC with tagged protons, focusing on the novel and interesting
information that the forward proton distributions can provide. We recall that such measurements are unlikely to be possible at other colliders
in the near future. 

The formalism used to calculate the perturbative  quarkonium CEP cross section is explained in detail
in \cite{HarlandLang09}~--~\cite{khrysthal}.
The expected cross sections and final-state particle distributions (in particular of the outgoing protons) are  determined by a non-trivial convolution of the hard amplitude $T$ and the so-called
 soft survival factors $S^2$, defining the probability that the
rapidity gaps survive soft and semi-hard rescattering effects (see~\cite{acta} for a review).
This is modelled in the SuperCHIC Monte Carlo~\cite{SuperCHIC}, which allows for an exact generation
on an event-by-event basis of the  distributions of the final-state central particles and outgoing protons, as well as a precise evaluation of the expected acceptances after experimental cuts have been imposed.

\begin{table}[ht]
\begin{tabular}{|l|c|c|c|c|c|}
\hline
$\sqrt{s}$ (TeV)&0.5&1.96&7&10&14\\
\hline
$\frac{{\rm d}\sigma}{{\rm d}y_{\chi_c}}(pp\to pp(J/\psi\,+\,\gamma))$&0.57&0.73&0.89&0.94&1.0\\
\hline
$\frac{{\rm d}\sigma(1^+)}{{\rm d}\sigma(0^+)}$&0.59&0.61&0.69&0.69&0.71\\
\hline
$\frac{{\rm d}\sigma(2^+)}{{\rm d}\sigma(0^+)}$&0.21&0.22&0.23&0.23&0.23\\
\hline
\end{tabular}
\caption{Differential cross section (in nb) at $y_\chi=0$ for $\chi_{cJ}$ CEP via the $\chi_{cJ} \to J/\psi\gamma$ decay, summed over the $J=0,1,2$, and cross section ratios of $\chi_{c(1,2)}$ to $\chi_{c0}$ production.}\label{chires1}
\end{table}

As  shown in~\cite{HarlandLang09,HarlandLang10}, the $\chi_{c(1,2)}$ CEP rates are expected to be heavily suppressed relative to the $\chi_{c0}$, due to the near-exact $J_z=0$ selection rule that operates for CEP~\cite{Khoze00a,kkmr}, although this suppression may be compensated by the larger $\chi_{c(1,2)}\to J/\psi \gamma$ branching ratios if $\chi_c$ CEP is observed via this decay channel (the $\chi_{c1}$($\chi_{c2}$) mesons have 30(17)$\times$ higher branching ratios respectively). In Table~\ref{chires1} we therefore show predictions \cite{HarlandLang10} for the $pp\to pp(\chi_c) \to pp(J/\psi\gamma)$ process at RHIC, Tevatron and LHC energies.
 We note (see also~\cite{teryaev}) that a significant fraction of the $\chi_c$ events are expected to correspond to the higher spin $\chi_{c(1,2)}$ states.
Unfortunately, the M($J/\psi+\gamma$) mass resolution in the CDF measurement \cite{cdfchic}
did not allow a clear separation of the $\chi_c$ states, and so this prediction could not be verified. However, it may be possible to isolate
the $\chi_{c0}$ CEP contribution via two-body decay channels, with the $\chi_c \to \pi\pi$ decay being a promising example.
 We recall that such hadronic channels, especially $\pi\pi$, $K^+K^-$ and $p\bar{p}$, are ideally suited for spin-parity analysis of the $\chi_c$ states: in particular the fact that the $\chi_{c(1,2)}$ two body branching ratios are in general of the same size or smaller (or even absent for the $\chi_{c1}$) than the $\chi_{c0}$ ensures that the $J_z=0$ selection rule is active, see \cite{HarlandLang09,Khoze04}. In the case of two-body and four-body channels the mass resolution
should be much better (of order of a few MeV in the case of STAR ~\cite{wlodek}) than in the previously observed $\chi_c\to J/\psi \gamma$ channel. 
\begin{table}[h]
\begin{tabular}{|l|c|c|c|c|}
\hline
$\sqrt{s}$ (TeV)&1.96&7&10&14\\
\hline
$\frac{{\rm d}\sigma}{{\rm d}y_{\chi_b}}(pp\to pp(\Upsilon\,+\,\gamma))$&0.56&0.70&0.73&0.74\\
\hline
$\frac{{\rm d}\sigma(1^+)}{{\rm d}\sigma(0^+)}$&0.029&0.032&0.032&0.034\\
\hline
$\frac{{\rm d}\sigma(2^+)}{{\rm d}\sigma(0^+)}$&0.077&0.081&0.081&0.083\\
\hline
\end{tabular}
\caption{Differential cross section (in pb) at  $y_\chi=0$ for $\chi_{bJ}$ CEP via the $\chi_{bJ} \to \Upsilon\gamma$ decay, summed over the $J=0,1,2$, and cross section ratios of $\chi_{b(1,2)}$ to $\chi_{b0}$ production.} \label{chires3}
\end{table}

Table~\ref{chires3} shows predictions \cite{HarlandLang10} for  the central exclusive $pp\to pp(\chi_b) \to pp(\Upsilon\gamma)$ process at Tevatron and LHC energies. While the overall rate is greatly reduced compared to $\chi_c$ production, $\chi_b$ CEP remains a potential observable at the LHC.
 We can see that $\chi_{b1}$ will give a negligible contribution to the overall rate, while the relative $\chi_{b2}/\chi_{b0}$ contribution is reduced in comparison to the $\chi_c$ case.

Finally, we show in Table~\ref{chires4} predictions 
for $\eta_c$ and $\eta_b$ CEP  at Tevatron and LHC energies \cite{HarlandLang10}.
In both cases, as a result of the $J_z^P=0^+$ selection rule the expected rates are roughly two orders of magnitude smaller
 than the associated $\chi_{c,b}$ cross sections.
We  also see that the cross sections are only slowly decreasing with energy.
In particular,
the $\chi_c$ and $\eta_c$ rates  at RHIC are not
significantly lower than the Tevatron predictions.
This is due to the survival factors ($S^2$) which increase with decreasing $\sqrt{s}$, and, thus,  compensate 
the decrease in the CEP cross section coming from the smaller gluon densities  at RHIC energies \cite{khrysthal}. 
\begin{table}[h]
\begin{tabular}{|l|c|c|c|c|}
\hline
$\sqrt{s}$ (TeV)&1.96&7&10&14\\
\hline
$\frac{{\rm d}\sigma}{{\rm d}y_\eta}(\eta_c)$&200&200&190&190\\
\hline
$\frac{{\rm d}\sigma}{{\rm d}y_\eta}(\eta_b)$&0.15&0.14&0.14&0.12\\
\hline
\end{tabular}
\caption{Differential cross section (in pb) at $y_{\eta}=0$ for  $\eta_{b,c}$ CEP.}\label{chires4}
\end{table}

In~\cite{HarlandLang10,kkmr},
it was shown that the distributions in $p_\perp$ and difference in azimuthal angle $\phi$ of the outgoing protons depend sensitively on the spin and parity of the centrally produced object. This was studied in detail in \cite{khrysthal} where 
various plots for the expected ${\rm d}\sigma/{\rm d}\phi$ distributions are given
for  CEP of $\chi_{c(0,1,2)}$ and $\eta_c$ states at $\sqrt{s}=500$ GeV at RHIC.
It is also demonstrated that, by
applying different cuts to the outgoing proton $p_\perp$, we can probe the underlying theory in a more detailed way. We observe that
for low $p_\perp$ the screening corrections do not affect the `bare' behaviour too much, 
while in the case of a relatively large $p_\perp$  the role of absorptive effects becomes quite visible:
starting from $\phi=0$ the absorptive correction increases with $\phi$ producing the diffractive dip structure in the region of $\phi\sim \pi/2$ for the cases of the $\chi_{c0}$ and $\chi_{c2}$ and about $\phi\sim 2.3$ for the $\chi_{c1}$. These characteristic `diffractive dip' structures have the same physical origin as the proton azimuthal distribution patterns first discussed in~\cite{KMRtag}. 
 
A further way to extract spin information about the centrally produced $\chi$ state is by measuring the angular distributions of its decays products, in particular, the final state $\mu^+\mu^-$ pair. These spin-dependent angular distributions  would also represent an interesting observable, providing complementary information to the tagged proton distributions. A potentially very promising 
measurement (especially with the STAR detector at RHIC)
is  
$\chi_{c0}$ CEP, via two-body
 (e.g. $\chi_{c0} \to \pi^+\pi^-,\, K^+K^-,\,p\overline{p}$) or four-body (e.g. $\chi_{c0} \to 2(\pi^+\pi^-),\,\pi^+\pi^-K^+K^-$)
channels, 
 provided the non-resonant QCD backgrounds are sufficiently under control (we recall that the CEP of
higher spin $\chi_{c(1,2)}$ states are expected to give negligible contributions via these decay channels).
 In the case of $\eta_c$ production, the three-body (e.g. $K\overline{K}\pi$), and four-body (e.g. direct $2(\pi^+\pi^-)$) decay modes appear to be quite promising.
The most important uncertainties that are present in our calculation are addressed in~\cite{HarlandLang09}~--~\cite{khrysthal}. 

 To conclude, the results of studies in \cite{HarlandLang09}~--~\cite{khrysthal} demonstrate the rich phenomenology that 
quarkonium CEP processes offer at high-energy colliders.
It is also worth mentioning that  CEP can help to shed light on the nature of the numerous recently discovered new charmonium-like mesons X,Y,Z~\cite{PDG}.

\section*{Acknowledgements}

We thank  Mike Albrow, Wlodek Guryn, Alan Martin, Risto Orava and Rainer Schicker 
for discussions. VAK thanks the Organizers for 
providing an excellent scientific  environment at the Workshop and for the support.


\begin{thebibliography}{9}

\bibitem{HarlandLang09}
L.~A.~Harland-Lang, V.~A.~Khoze, M.~G.~Ryskin and W.~J.~Stirling, \emph{Eur.\ 
Phys.\ J.}  {\bf C65}, 433 (2010).

\bibitem{HarlandLang10}
L.~A.~Harland-Lang, V.~A.~Khoze, M.~G.~Ryskin and W.~J.~Stirling,
arXiv:1005.0695 [hep-ph].

\bibitem{khrysthal}
L.~A.~Harland-Lang, V.~A.~Khoze, M.~G.~Ryskin and W.~J.~Stirling,
arXiv:1011.0680 [hep-ph].

\bibitem{KMRprosp} V.~A.~Khoze, A.~D.~Martin and M.~G.~Ryskin,
\emph{Eur.\ Phys.\ J.}  {\bf C23}, 311 (2002).

\bibitem{acf} M.G.~Albrow, T.D.~Coughlin and J.R.~Forshaw,
arXiv:1006.1289;\\ 
\emph{Prog. Part. Nucl. Phys.} (to be published).

\bibitem{MA}M.~Albrow, this workshop (DIFF2010).

\bibitem{wlodek} W.~Guryn, this workshop (DIFF2010).

\bibitem{schciker} R.~Schicker, this workshop (DIFF2010).

\bibitem{Royon} C.~Royon, this workshop (DIFF2010).

\bibitem{Khoze04}
V.~A.~Khoze, A.~D.~Martin, M.~G.~Ryskin and W.~J.~Stirling, \emph{Eur.\ Phys.\ 
J.}  {\bf C35}, 211 (2004).

\bibitem{HKRSTW} S.~Heinemeyer {\it et al.}, \emph{Eur.\ Phys.\ J.}  {\bf C53}, 
231 (2008).

\bibitem{Khoze04gg}
V.~A.~Khoze, A.~D.~Martin, M.~G.~Ryskin and W.~J.~Stirling, \emph{Eur.\ Phys. 
J.}  {\bf C38} (2005) 475.

\bibitem{teryaev} R.~S.~Pasechnik, A.~Szczurek and O.~V.~Teryaev,
\emph{Phys.\ Lett.} {\bf B680}, 62 (2009).

\bibitem{Khoze00a}
V.~A.~Khoze, A.~D.~Martin and M.~G.~Ryskin, \emph{Eur.\ Phys.\ J.}  {\bf C19}, 
477 (2001) [\emph{ibid.}  {\bf C20}, 599 (2001)].

\bibitem{kkmr} A.B.Kaidalov, V.A.Khoze, A.D.Martin, and M.G.Ryskin, \emph{Eur. 
Phys. J.} {\bf C31} 387 (2003).

\bibitem{ways} A.De Roeck {\it et al.}, \emph{Eur. Phys. J.} {\bf C25}, 391 (2002).

\bibitem{fp420} M.G.Albrow {\it et al.}, 
\emph{J. Inst.} {\bf 4} T10001 (2009).

\bibitem{cdfchic}
T.~Aaltonen {\it et al.}  [CDF Collaboration],
\emph{Phys.\ Rev.\ Lett.}  {\bf 102}, 242001 (2009).

\bibitem{RO} R. Orava, this workshop (DIFF2010).

\bibitem{fsc} M.G.~Albrow {\it et al.}, \emph{JINST} {\bf 4} P10001 (2009);
J.~W.~Lamsa and R.~Orava, \emph{JINST} {\bf 4}, P11019 (2009).

\bibitem{acta} A.~D.~Martin, M.~G.~Ryskin and V.~A.~Khoze,
\emph{Acta Phys.\ Polon.}  {\bf B40}, 1841 (2009).

\bibitem{SuperCHIC} The MC code and documentation are available at {\tt http://projects.hepforge.org/superchic/}.

\bibitem{KMRtag} V.~A.~Khoze, A.~D.~Martin and M.~G.~Ryskin,
\emph{Eur.\ Phys.\ J.}  {\bf C24}, 581 (2002).

\bibitem{PDG} K. Nakamura {\it et al.} (Particle Data Group), 
\emph{Journal of Physics} {\bf G37}, 075021 (2010).

\end{thebibliography}
\end{document}